\documentclass[aps,10pt,pra,twocolumn,showpacs,showkeys,groupedaddress]{revtex4-1}
\usepackage[colorlinks,linkcolor=blue,citecolor=blue,filecolor=cyan,urlcolor=blue]{hyperref}
\usepackage{amsmath}
\usepackage{graphicx}
\usepackage{graphicx}			% for inputting .EPS figures
\usepackage{bm}
\usepackage{xcolor}
\usepackage{comment}
\usepackage[normalem]{ulem}

\DeclareBoldMathCommand{\bfmu}{\mu}
\newcommand{\be}{\begin{equation}}
\newcommand{\ee}{\end{equation}}
\newcommand{\bea}{\begin{eqnarray}}
\newcommand{\eea}{\end{eqnarray}}
\def\mum{\rm\mu m}

\def\eg{{\it e.g.,\/}}
\def\vs{{\it vs.\/}}

\def\GP{Gross-Pitaevskii}

\begin{document}

\title{Long spatial coherence times a few micro-meters from a room temperature surface}

\author{Shuyu Zhou}	
	\affiliation{Department of Physics, Ben-Gurion University of the Negev, Be'er Sheva 84105, Israel}
%\author{Julien Chab\'e}
%	\altaffiliation{Present address: Observatoire de la C\^ote d'Azur, Universit\'e de Nice-Sophia Antipolis, CNRS, Parc Valrose, F-06108 Nice Cedex~2, France.}
%\author{Ran Salem}
%	\altaffiliation{Present address: Physics Department, Nuclear Research Center Negev, Beer-Sheva 84109, Israel.}
%	\affiliation{Department of Physics, Ben-Gurion University of the Negev, Be'er Sheva 84105, Israel}
%\author{Tal David}
%	\affiliation{Department of Physics, Ben-Gurion University of the Negev, Be'er Sheva 84105, Israel}
\author{David Groswasser}
	\affiliation{Department of Physics, Ben-Gurion University of the Negev, Be'er Sheva 84105, Israel}	
\author{Mark Keil}
	\thanks{Corresponding author}
	\email{\tt mkeil@netvision.net.il}
	\affiliation{Department of Physics, Ben-Gurion University of the Negev, Be'er Sheva 84105, Israel}	
\author{Yonathan Japha}
	\affiliation{Department of Physics, Ben-Gurion University of the Negev, Be'er Sheva 84105, Israel}	
\author{Ron Folman}
	\affiliation{Department of Physics, Ben-Gurion University of the Negev, Be'er Sheva 84105, Israel}

%\begin{comment}
\begin{abstract}
The search for quantum coherence based on isolated atoms integrated with a room temperature solid state device (so-called atomchip~\cite{AdvAtMol48-263,APB74-469,RevModPhys79-235}) has been intensifying in the last decade, with advances being made towards applications such as clocks, quantum information processing, surface probing and acceleration and gravitational field sensors. Such a device will also enable (and to some extent has already enabled) novel experiments in fundamental physics (\eg~\cite{PRL98-063201,PRL105-080403,Nature464-1170,NaturePhysics9-640}). Here we report on the trapping and maintenance of spatial coherence of atoms (in a Bose-Einstein Condensate~--~BEC) about~$\rm5\,\mum$ from a room temperature surface, reducing significantly the distance previously achieved between the spatially coherent atoms and their classical 
environment~\cite{PRL94-090405,NaturePhysics1-57,PRL98-030407,NaturePhysics5-592,PRL105-243003}, and most importantly entering the regime where atomic circuits are enabled. In addition, we enter the interesting regime in which the distance to the surface is much smaller than the probed coherence length, a regime in which the spatial dephasing reaches its maximal rate.
\end{abstract}
%\end{comment}

\date{\today}

%?? \pacs{03.65.Wj, 03.75.-b, 37.10.Gh, 67.85.-d} removed for ArXiv

%?? \keywords{keywords go here} removed for ArXiv

\maketitle

To realize a ``solid state'' device with isolated atoms, at least three milestones~--~adapted from electronic devices~--~need to be met: arbitrary (\eg\ not periodic) guides and traps, single site addressability, and controlled interaction via tunneling barriers. The latter requirement demands that potentials must be sculptured with a resolution on the scale of the de-Broglie wavelength of the atoms (about~$\rm1\,\mum$). To achieve these milestones in a scalable device, thereby forming circuits for matter-waves (\eg~\cite{PRL103-140405,PRA74-012308}), one must be able to trap the atoms and manipulate them coherently a few~$\mum$ or less from the surface used to generate the potential fields~\cite{NewJPhys12-023039}. An interference or diffraction pattern, the hallmark of spatial coherence, from trapped atoms close to the surface, has so far not been observed. 

We study the spatial coherence by loading the~BEC into a lattice potential and observing diffraction. While lattices with trapped atoms close to surfaces have been realized~\cite{RevSciInst85-053102,PRA89-051602} and while diffraction has been observed from atoms dynamically reflected from surfaces (see~\cite{PRX4-011029} and references therein), this is the first time a~BEC is trapped in a lattice close to the surface and then allowed to evolve into a diffraction pattern proving that its coherence length spans at least several lattice sites. 

We provide new experimental insight regarding the interplay between the rate in which surface Johnson noise affects the spatial coherence length of a~BEC~\cite{ApplPhysB76-173,PRA69-043602} and the rate in which a~BEC can phase-lock itself. In addition, this work shows that potential corrugations due to material and fabrication impurities~\cite{PRL92-076802,PRB77-201407(R),Science319-1226} may be reduced to a non-inhibiting level.  This work may enable studies of low dimensional gases (\eg\ Tonks gas) with a single sample, many-body rephasing rate against external small-correlation-length noise, ultra-sensitive probing of surface effects such as the Casimir-Polder force and the hypothesized short-range fifth force, as well as devices with matter-waves such as acceleration sensors based on counter-propagating~3D traps moving on a loop (Sagnac)~\cite{PRL98-030407,PRA75-063406}.

\begin{figure}[t!]
   \vskip-\baselineskip
      \centering
      \includegraphics[width=0.5\textwidth]{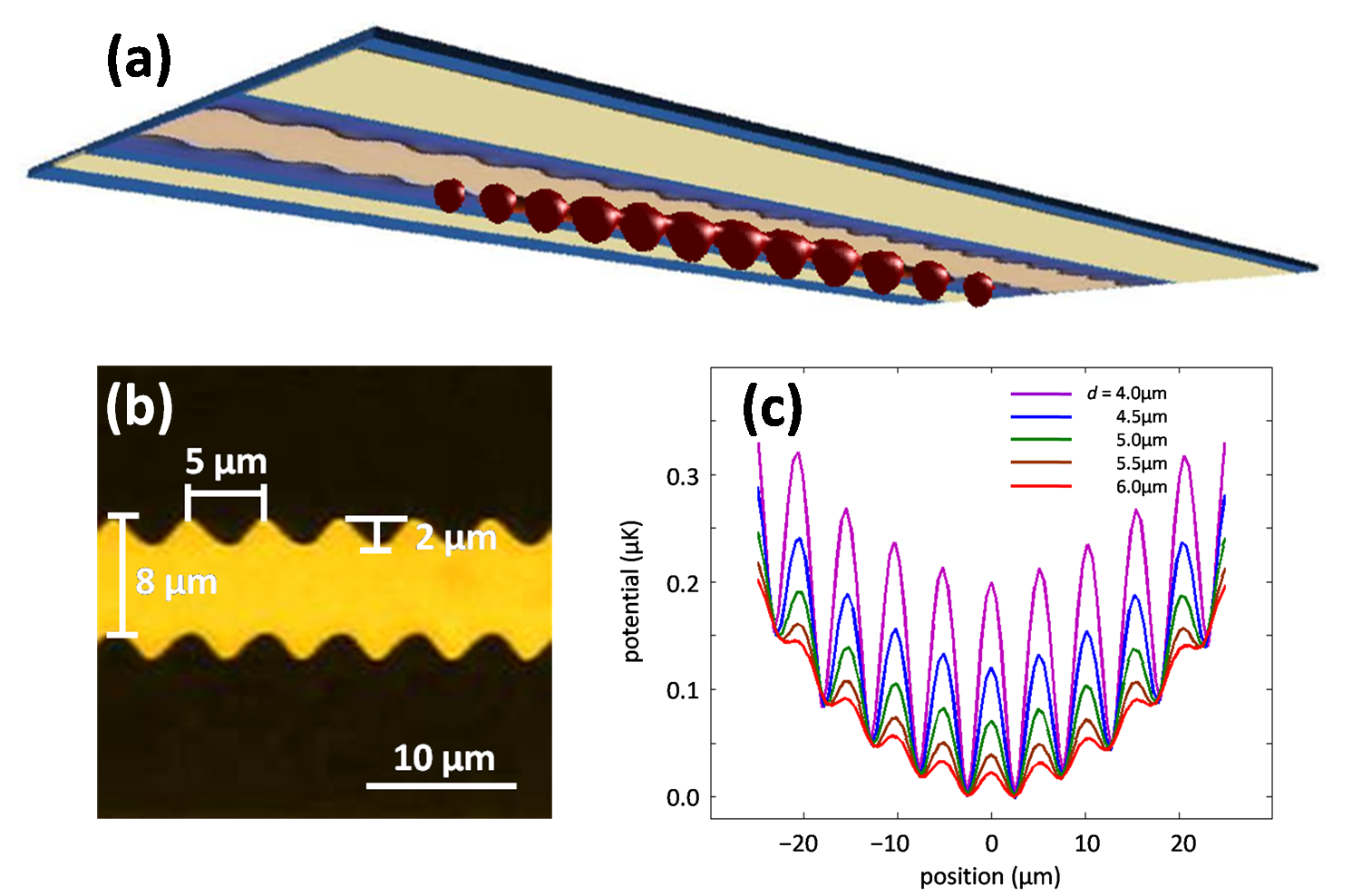}
   \vskip-0.5\baselineskip
   \caption{(Color online) Experimental configuration. (a)~An artist's view of the trapped cloud of atoms a few~$\mum$ from the surface. The atoms are trapped below the surface to allow their state to evolve after release from the trap without falling onto the chip. (b)~An optical image of the current-carrying snake wire creating the~1D lattice potential. It adds a magnetic modulation onto the main trapping potential created by a straight trapping wire (not shown). The snake wire is made of gold and is~$\rm500\,nm$ thick. Its other dimensions are shown in the image. (c)~Using the trapping wire to adjust the height~$d$ of the trap from the snake wire strongly affects the potential modulation, shown here with the~$\rm5\,\mum$ periodicity acquired from the wire at constant current (adjusting the current allows to fine-tune the barrier amplitude). The modulated potential is depicted here together with a weak harmonic potential produced by the trapping wire, giving rise to the atom density profile shown in~(a). Not shown is the radial confinement potential that prevents the atoms from hitting the surface. We are not able to provide {\it in-situ} images of the modulated atom density as our imaging resolution is~$\sim\rm7\,\mum$.}
   \label{fig:schematic}
\end{figure}

The experiment is conducted as follows: we create a~BEC (typically~$10^4$ atoms in the~F=2,~m$\rm_F$=2 state) and load a magnetic trap located about~$\rm5\,\mum$ from the conductive surface of an atom chip. Different from previous trapped atom interferometers, we utilize only~DC magnetic fields. The trap potential is modulated to create a~1D lattice. This~$\rm5\,\mum$ modulation is due to a meandering wire (``snake wire'') which causes the electron current to periodically change direction thus creating magnetic barriers. The confining potential is mostly due to a larger wire (``trapping wire'') underneath the snake wire and fields from external coils. As shown in~Fig.~\ref{fig:schematic}, the modulation of the potential may be controlled by the height of the trap from the surface (as well as the snake wire current). Indeed, in the experiment when the height is made a few~$\mum$ larger, no diffraction pattern is observed, and when the height is made a few~$\mum$ smaller, higher diffraction orders are observed. Aside from the~1D lattice potential, longitudinal confinement via a harmonic potential is kept on throughout the trapping. As we work in the regime of a few hundred atoms per~$\mum$, we increase the field at the trap minimum to~$\rm18.3\,G$ in order to decrease the trap frequency ratio (longitudinal to transverse) so as not to enter the~1D regime in which the initial coherence length drops drastically. Based on {\it in-situ} imaging, the~BEC length covers about~6-8 lattice sites. 

We hold the atoms in this trap for periods up to~$\rm500\,ms$ and then release them. The release includes two steps: in the first~$\rm2.3\,ms$ we push the atoms away from the surface (``launching'') by increasing the current in the snake wire from~$\rm5.5\,mA$ to~$\rm18\,mA$. During this step the cloud goes through parabolic acceleration by the longitudinal potential creating a focus after full release. After that, all potentials are turned off and the cloud experiences another~$\rm11.7\,ms$ of free-fall under gravity (time-of-flight:~TOF), following which an absorption image is taken. The focusing plays a crucial role in the experiment; the diffraction pattern periodicity of~$\rm15\,\mum$ would not have, under normal free-fall, produced any observable visibility, as the BEC length, and therefore the expected width of the diffraction peaks, is bigger than~$\rm30\,\mum$.

\begin{figure}[t!]
   \centering 
   \includegraphics[width=0.45\textwidth]{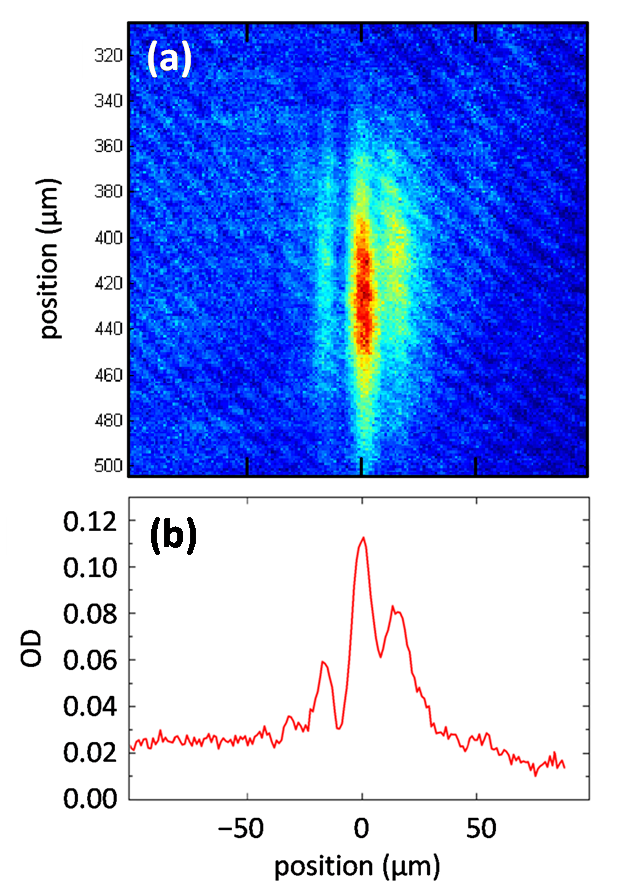}
   \caption{(Color online) Experimental signal. Diffraction patterns observed after a trap holding time of~$t=\rm100\,ms$ for~30 consecutive experimental cycles (no post correction or post selection is used). (a)~Average of images acquired by absorption imaging. (b)~A cut through the center of~(a) showing the optical density~(OD). The high visibility is typical of the superfluid phase~\cite{Nature415-39} or Josephson regime~\cite{JPhysB40-R61}. We have been able to reproduce the observed asymmetry in the amplitude of the first-order diffraction peaks by introducing a linear phase difference gradient between the sites~($\rm1\,rad$ increase between two adjacent sites). Similarly the asymmetry may be due to periodic imperfections in the fabrication process~\cite{IntJOpt17-857}.}   
   \label{fig:experiment}
\end{figure}

An average of~30 consecutive shots is presented in Fig.~\ref{fig:experiment}a-b ($\rm100\,ms$ holding time). The zero-order and the first-order diffraction peaks are clearly visible. We have confirmed that the observed diffraction pattern periodicity of~$\rm15\,\mum$ is independent of the height of the trap, the purity of the~BEC, the position of the trapped cloud along the lattice and the amplitude of the diffraction orders.

\begin{figure}[t!]
   \centering 
   \includegraphics[width=0.45\textwidth]{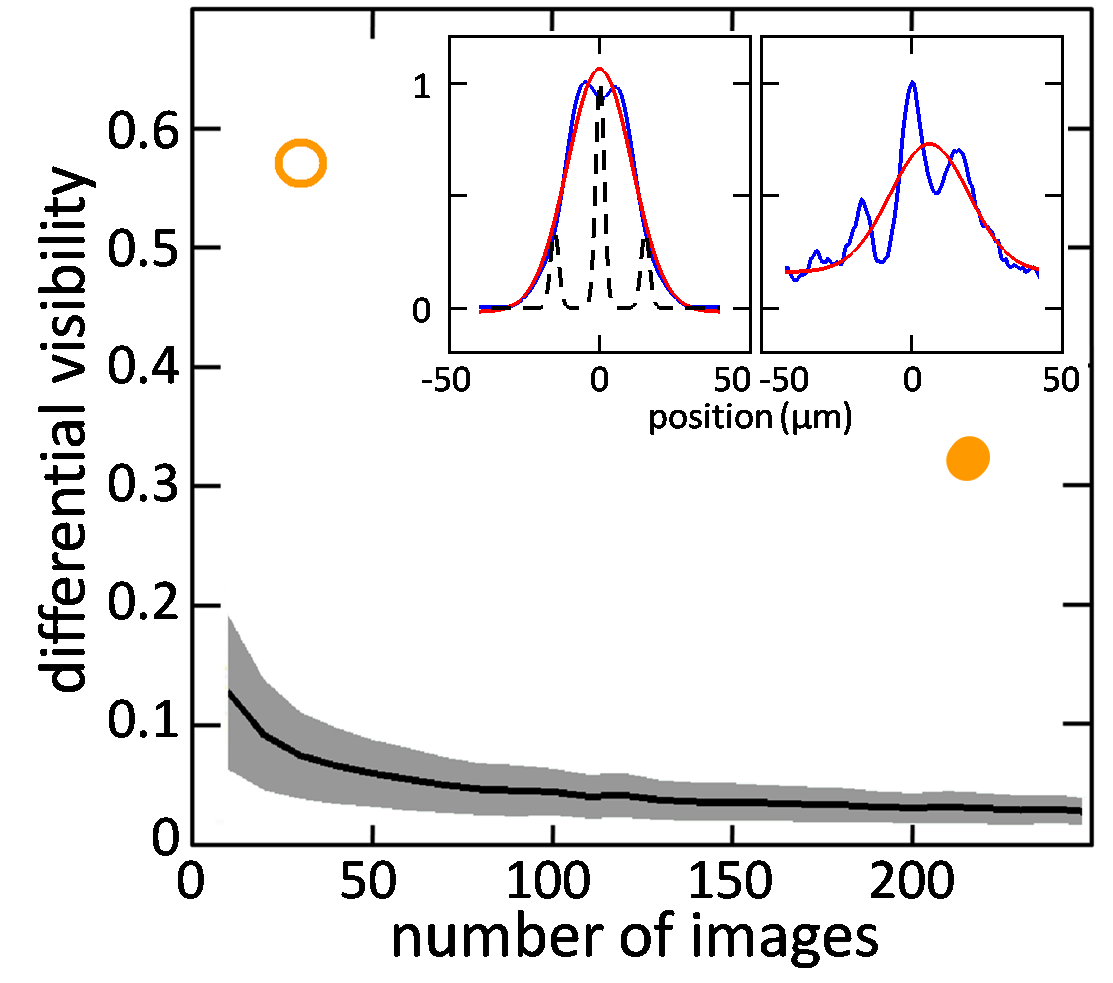}
   \caption{(Color online) Visibility. A comparative analysis (Fig.~\ref{fig:experiment}b \vs\ simulation with random phases) to show that the observed signal (open circle) may not come from a random phase (black line). The grey area describes the simulated standard deviation from~500 runs. The compared variable (differential visibility) is simply the maximal deviation from the best Gaussian fit. (The standard fits of a convolution of a sine function with a Gaussian or three Gaussian peaks with varying amplitude and width, proved to be too unstable). An additional data point is given for a sample of~215 experimental shots (full circle) and is taken from the data set described in Fig~\ref{fig:coherence}a. The insets describe the fit to a Gaussian for the data of Fig.~\ref{fig:experiment}b (right) and an average of~30 shots with random phases (left). The dashed black line is the result of the simulation if the phases in the different sites are made to be equal.}   
   \label{fig:visibility}
\end{figure}

The rather surprising result of this experiment is that beyond the known fact that a~BEC survives for a significant time so close to a room temperature surface, it also maintains its spatial coherence length, thus enabling atomic circuits. As a split~BEC is expected to give rise to interference even when only short range spatial coherence exists (\eg\ as in the free evolution of a~1D~BEC), we present in Fig.~\ref{fig:visibility} a comparative analysis (experiment and simulation) to show that the observed signal may not come from a random phase. The simulation creates a diffraction pattern evolving from a state in which the phases between the different lattice sites are random. As the position of the peaks is different for each shot originating from a random phase distribution, an average of such shots exhibits low visibility. The compared variable shows a significant number of standard deviations from the experimental signal, thus proving the experimental signal is due to coherent diffraction, which in turn indicates that the coherence length covers at least several lattice sites.

\begin{figure}[t!]
      \includegraphics[width=0.45\textwidth]{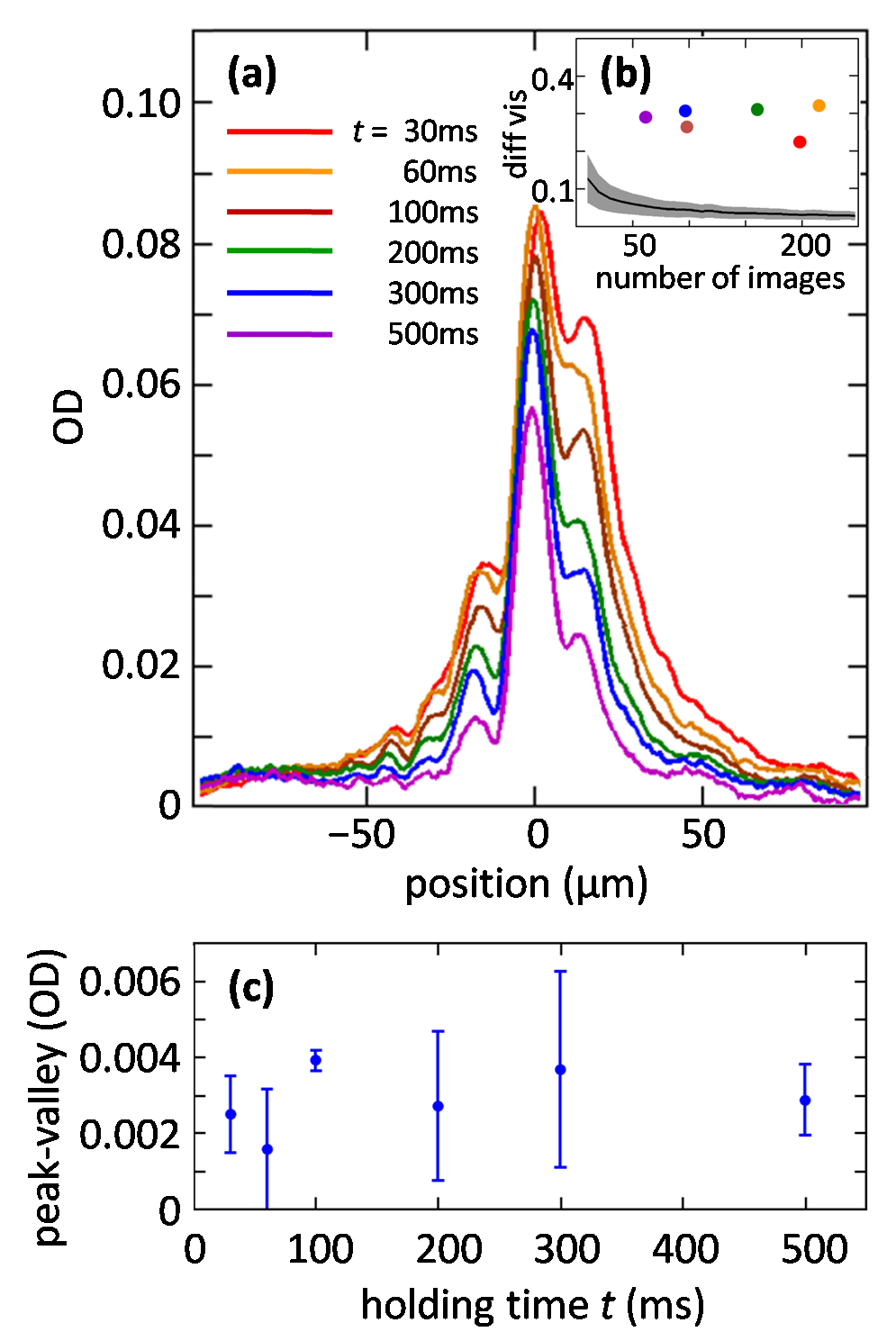}
      \vskip-0.5\baselineskip
   \caption{(Color online) Spatial coherence. (a)~Repeating the cut of Fig.~\ref{fig:experiment}b (this time fully integrating the signal along the vertical axis) for trap holding times of~30-$\rm500\,ms$, averaged over~100-200 experimental cycles for each holding time. The data are taken for trap heights~$5<d<\rm6.2\,\mum$. Some contrast is lost relative to Fig.~\ref{fig:experiment}b due to the much larger number of experimental cycles used, together with slight drifts of the lattice and experimental conditions during these measurements. The overall~OD decreases with time due to the finite lifetime of the cloud. Inset: (b)~The same as Fig.~\ref{fig:visibility} for the data sets of~(a)~--~same color coding as~(a). (c)~Un-normalized visibility of the first order peaks and valleys as a measure of coherence \vs\ trap holding time. The error bars are taken from the difference between the~$-1$ and~$+1$ order visibilities. The data
is consistent with a flat line which results if~$\Gamma_{\rm D}/\Gamma_{\rm R}<1$~\cite{Gamma_ratio}.}
%is consistent with a flat line which results if~$\Gamma_{\rm D}/\Gamma_{\rm R}<1$\footnote{$\Gamma_{\rm R}$  is related to the Josephson frequency ($\omega_{\rm J}$). In principle $\Gamma_{\rm R}\simeq\omega_{\rm J}/10$~\cite{PRA80-053619} and~$\Gamma_{\rm D}$ is the same as the spin-flip rate originating from short-correlation-length noise~\cite{ApplPhysB76-173}. The total spin-flip rate in our trap is~$\rm3\,Hz$. As for our specific experiment we have significant uncertainties concerning the values of the chemical potential, the tunneling rate and the correlation length of the technical noise, we do not quote here an estimated value for~$\Gamma_{\rm D}/\Gamma_{\rm R}$.}.} 
   \label{fig:coherence}
\end{figure}

To quantify the spatial coherence dephasing rate, in Fig.~\ref{fig:coherence}a we present the averaged picture of a sample of about~1000 pictures at different holding times. As the data taking of this large sample spanned numerous days and changing experimental conditions, slight drifts in the experiment had to be accounted for. Consequently, in contrast to the result presented in Fig.~\ref{fig:experiment}a-b, the data analysis in Fig.~\ref{fig:coherence}a includes post-selection depending on height and horizontal position of the trapped cloud. \linebreak {\it In-situ} images every few experimental cycles provide the required position and height information. In addition, we have moved the position of the fringe pattern by the same shift observed in the position of the trapped cloud. Fig.~\ref{fig:coherence}b is a repeat of Fig.~\ref{fig:visibility}, for the data sets of Fig.~\ref{fig:coherence}a.

In Fig.~\ref{fig:coherence}c we present the coherence as a function of holding time. We estimate the relative coherence for the different holding times directly from the visibility, assuming that imperfections in the experiment and data analysis which lower the visibility (\eg\ relative to Fig.~\ref{fig:experiment}b) are independent of the holding time. The data is consistent with no dephasing and consequently one can with caution estimate for our experiment an upper bound of about one on the ratio~$\Gamma_{\rm D}/\Gamma_{\rm R}$ of the spatial dephasing rate due to Johnson noise over the spatial rephasing rate of the~BEC. This is rather striking when one recalls that the~BEC length is about~$\rm30\,\mum$ and the correlation length of the Johnson noise at a height of~$\rm5\,\mum$ is~$\rm5\,\mum$~\cite{ApplPhysB76-173}. In this regime where spatial coherence is probed on a much larger scale than the distance to the surface, the dephasing rate is expected to be maximal~\cite{ApplPhysB76-173}. Let us note that technical noise should have no effect on the spatial coherence length as it typically has a large correlation length (not so for shot noise). Technical noise may introduce secondary effects through heating and spin-flips. Extrapolating to smaller heights and wires, our results show that atomic circuits are not inhibited by lack of spatial coherence due to the nearby surface. 

As an outlook, let us note that beyond the extensive~\GP\ simulations we have conducted, further theoretical work is required in order to better explain the specific details of the observed diffraction pattern. As for atomic circuits, the next step should be to exhibit a high level of control over the tunneling rates. Crossing the border to~$\Gamma_{\rm D}/\Gamma_{\rm R}>1$ by increasing the Johnson noise (\eg\ smaller heights) will further improve our theoretical understanding. The effect of Johnson noise on~BECs in guides is expected to be similarly weak and will enable atomic circuits with guides (\eg~\cite{PRL99-060402}), having the advantage of less phase diffusion due to atom-atom interactions. Advanced fabrication will contribute to improved results. For example, utilizing electrically anisotropic materials~\cite{EurPhysJD48-321} is expected to reduce even further the Johnson noise responsible for dephasing. Utilizing crystalline materials such as carbon nano-tubes~\cite{PRA79-043403} or graphene sheets is expected to reduce both Johnson noise and potential corrugations due to electron scattering. Finally, to reduce the atom-surface distance by another order of magnitude, thus improving control over tunneling barriers, it would be beneficial to use nano-wires in order to considerably increase the gradients so that they can overcome the Casimir-Polder potential, as suggested in our work~\cite{NewJPhys12-023039,PRA79-043403}. Small wires can also enable geometries which will allow more flexibility in the choice of the magnetic field at the trap bottom and the frequency ratio (longitudinal to transverse) in the trap.

\begin{acknowledgments}\vskip-\baselineskip

We thank Julien Chab\'e, Tal David and Ran Salem for the initial steps of the experiment, Zina Binstock for the electronics, and the~BGU nano-fabrication facility for providing the high-quality chip. We are grateful to Amichay Vardi and Carsten Henkel for helpful discussions. This work is funded in part by the Israeli Science Foundation, the~EC ``matter-waves'' consortium, and the German-Israeli~DIP project supported by the~DFG.

\end{acknowledgments}

\bibliography{spatial_coherence}

\begin{thebibliography}{10}

\bibitem{AdvAtMol48-263}
R.~Folman, P.~Kr\"uger, J.~Schmiedmayer, J.~Denschlag, and C.~Henkel.
\newblock {Microscopic atom optics: from wires to an atom chip}.
\newblock {\em Adv. At. Mol. Opt. Phys.}, 48:263, 2002.
\newblock
  \href{http://www.sciencedirect.com/science/article/pii/S1049250X02800118}
  {\tt doi=10.1016/S1049-250X(02)80011-8}.

\bibitem{APB74-469}
J.~Reichel.
\newblock {Microchip traps and Bose-Einstein condensation}.
\newblock {\em Appl. Phys. B}, 74:469, 2002.
\newblock \href{http://link.springer.com/article/10.1007/s003400200861}{\tt
  doi=10.1007/s003400200861}.

\bibitem{RevModPhys79-235}
J.~Fort\'agh and C.~Zimmermann.
\newblock {Magnetic microtraps for ultracold atoms}.
\newblock {\em Rev. Mod. Phys.}, 79:235, 2007.
\newblock
  \href{http://journals.aps.org/rmp/abstract/10.1103/RevModPhys.79.235}{\tt
  doi= 10.1103/RevModPhys.79.235}.

\bibitem{PRL98-063201}
J.~M. Obrecht, R.~J. Wild, M.~Antezza, L.~P. Pitaevskii, S.~Stringari, and
  E.~A. Cornell.
\newblock {Microscopic atom optics: from wires to an atom chip}.
\newblock {\em Phys. Rev. Lett.}, 98:063201, 2007.
\newblock
  \href{http://journals.aps.org/prl/pdf/10.1103/PhysRevLett.98.063201}{\tt
  doi=10.1103/PhysRevLett.98.063201}.

\bibitem{PRL105-080403}
K.~Maussang, G.~E. Marti, T.~Schneider, P.~Treutlein, Y.~Li, A.~Sinatra,
  R.~Long, J.~Est\`{e}ve, and J.~Reichel.
\newblock {Enhanced and reduced atom number fluctuations in a BEC splitter}.
\newblock {\em Phys. Rev. Lett.}, 105:080403, 2010.
\newblock
  \href{http://journals.aps.org/prl/pdf/10.1103/PhysRevLett.105.080403}{\tt
  doi=10.1103/PhysRevLett.105.080403}.

\bibitem{Nature464-1170}
M.~F. Riedel, P.~B\"ohi, Y.~Li, T.~W. H\"ansch, A.~Sinatra, and P.~Treutlein.
\newblock {Atom-chip-based generation of entanglement for quantum metrology}.
\newblock {\em Nature}, 464:1170, 2010.
\newblock
  \href{http://www.nature.com/nature/journal/v464/n7292/pdf/nature08988.pdf}{\tt
  doi=10.1038/nature08988}.

\bibitem{NaturePhysics9-640}
T.~Langen, R.~Geiger, M.~Kuhnert, B.~Rauer, and J.~Schmiedmayer.
\newblock {Local emergence of thermal correlations in an isolated quantum
  many-body system}.
\newblock {\em Nature Physics}, 9:640, 2013.
\newblock
  \href{http://www.nature.com/nphys/journal/v9/n10/pdf/nphys2739.pdf}{\tt
  doi=10.1038/nature08988}.

\bibitem{PRL94-090405}
Y.~J. Wang, D.~Z. Anderson, V.~M. Bright, E.~A. Cornell, Q.~Diot, T.~Kishimoto,
  M.~Prentiss, R.~A. Saravanan, S.~R. Segal, and S.~Wu.
\newblock {Atom Michelson Interferometer on a Chip Using a Bose-Einstein
  Condensate}.
\newblock {\em Phys. Rev. Lett.}, 94:90405, 2005.
\newblock
  \href{http://journals.aps.org/prl/pdf/10.1103/PhysRevLett.94.090405}{\tt
  doi=10.1103/PhysRevLett.94.090405}.

\bibitem{NaturePhysics1-57}
T.~Schumm, S.~Hofferberth, L.~M. Andersson, S.~Wildermuth, S.~Groth,
  I.~Bar-Joseph, J.~Schmiedmayer, and P.~Kr\"uger.
\newblock {Matter-wave interferometry in a double well on an atom chip}.
\newblock {\em Nature Physics}, 1:57, 2005.
\newblock \href{http://www.nature.com/nphys/journal/v1/n1/pdf/nphys125.pdf}{\tt
  doi=10.1038/nphys125}.

\bibitem{PRL98-030407}
G.-B. Jo, Y.~Shin, S.~Will, T.~A. Pasquini, M.~Saba, W.~Ketterle, D.~E.
  Pritchard, M.~Vengalattore, and M.~Prentiss.
\newblock {Long Phase Coherence Time and Number Squeezing of Two Bose-Einstein
  Condensates on an Atom Chip}.
\newblock {\em Phys. Rev. Lett.}, 98:030407, 2007.
\newblock
  \href{http://journals.aps.org/prl/pdf/10.1103/PhysRevLett.98.030407}{\tt
  doi=10.1103/PhysRevLett.98.030407}.

\bibitem{NaturePhysics5-592}
P.~B\"ohi, M.~F. Riedel, J.~Hoffrogge, J.~Reichel, T.~W. H\"ansch, and
  P.~Treutlein.
\newblock {Coherent manipulation of Bose-Einstein condensates with
  state-dependent microwave potentials on an atom chip}.
\newblock {\em Nature Physics}, 5:592, 2009.
\newblock
  \href{http://www.nature.com/nphys/journal/v5/n8/abs/nphys1329.html}{\tt
  doi=10.1038/nphys1329}.

\bibitem{PRL105-243003}
F.~Baumg\"artner, R.~J. Sewell, S.~Eriksson, I.~Llorente-Garcia, J.~Dingjan,
  J.~P. Cotter, and E.~A. Hinds.
\newblock {Measuring Energy Differences by BEC Interferometry on a Chip}.
\newblock {\em Phys. Rev. Lett.}, 105:243003, 2010.
\newblock
  \href{http://journals.aps.org/prl/pdf/10.1103/PhysRevLett.105.243003}{\tt
  doi=10.1103/PhysRevLett.105.243003}.

\bibitem{PRL103-140405}
R.~A. Pepino, J.~Cooper, D.~Z. Anderson, and M.~J. Holland.
\newblock {Atomtronic Circuits of Diodes and Transistors}.
\newblock {\em Phys. Rev. Lett.}, 103:140405, 2009.
\newblock
  \href{http://journals.aps.org/prl/pdf/10.1103/PhysRevLett.103.140405}{\tt
  doi=10.1103/PhysRevLett.103.140405}.

\bibitem{PRA74-012308}
E.~Charron, M.~A. Cirone, A.~Negretti, J.~Schmiedmayer, and T.~Calarco.
\newblock {Theoretical analysis of a realistic atom-chip quantum gate}.
\newblock {\em Phys. Rev. A}, 74:012308, 2006.
\newblock \href{http://journals.aps.org/pra/pdf/10.1103/PhysRevA.74.012308}{\tt
  doi=10.1103/PhysRevA.74.012308}.

\bibitem{NewJPhys12-023039}
R.~Salem, Y.~Japha, J.~Chab\'e, B.~Hadad, M.~Keil, K.~A. Milton, and R.~Folman.
\newblock {Nanowire atomchip traps for sub-micron atomâ€“surface
  distances}.
\newblock {\em New J. Phys}, 12:023039, 2010.
\newblock
  \href{http://iopscience.iop.org/1367-2630/12/2/023039/pdf/1367-2630_12_2_023039.pdf}{\tt
  doi=10.1088/1367-2630/12/2/023039}.

\bibitem{RevSciInst85-053102}
V.~Y.~F. Leung, D.~R.~M. Pijn, H.~Schlatter, L.~Torralbo-Campo, A.~L.~La Rooij,
  G.~B. Mulder, J.~Naber, M.~L. Soudijn, A.~Tauschinsky, C.~Abarbanel,
  B.~Hadad, E.~Golan, R.~Folman, and R.~J.~C. Spreeuw.
\newblock {Magnetic-film atom chip with 10$\mu$m period lattices of microtraps
  for quantum information science with Rydberg atoms}.
\newblock {\em Rev. Sci. Inst.}, 85:053102, 2014.
\newblock
  \href{http://scitation.aip.org/content/aip/journal/rsi/85/5/10.1063/1.4874005}{\tt
  doi=10.1063/1.4874005}.

\bibitem{PRA89-051602}
S.~Jose, P.~Surendran, Y.~Wang, I.~Herrera, L.~Krzemien, S.~Whitlock,
  R.~McLean, A.~Sidorov, and P.~Hannaford.
\newblock {Periodic array of Bose-Einstein condensates in a magnetic lattice}.
\newblock {\em Phys. Rev. A}, 89:051602, 2014.
\newblock \href{http://journals.aps.org/pra/pdf/10.1103/PhysRevA.89.051602}{\tt
  doi= 10.1103/PhysRevA.89.051602}; see also arXiv:
  {\href{http://arxiv.org/abs/1411.4730}{\tt doi= 1411.4730} (2015)}.

\bibitem{PRX4-011029}
H.~Bender, C.~Stehle, C.~Zimmermann, S.~Slama, J.~Fiedler, S.~Scheel, S.~Y.
  Buhmann, and V.~N. Marachevsky.
\newblock {Probing Atom-Surface Interactions by Diffraction of Bose-Einstein
  Condensates}.
\newblock {\em Phys. Rev. X}, 4:011029, 2014.
\newblock \href{journals.aps.org/prx/pdf/10.1103/PhysRevX.4.011029}{\tt
  doi=10.1103/PhysRevX.4.011029}.

\bibitem{ApplPhysB76-173}
C.~Henkel, P.~Kr\"uger, R.~Folman, and J.~Schmiedmayer.
\newblock {Fundamental limits for coherent manipulation on atom chips}.
\newblock {\em Appl. Phys. B}, 76:173, 2003.
\newblock
  \href{http://link.springer.com/article/10.1007%2Fs00340-003-1112-z#page-1}{\tt
  doi= 10.1007/s00340-003-1112-z}.

\bibitem{PRA69-043602}
C.~Henkel and S.~A. Gardiner.
\newblock {Decoherence of Bose-Einstein condensates in microtraps}.
\newblock {\em Phys. Rev. A}, 69:043602, 2004.
\newblock \href{http://journals.aps.org/pra/pdf/10.1103/PhysRevA.69.043602}{\tt
  doi= 10.1103/PhysRevA.69.043602}.

\bibitem{PRL92-076802}
D.-W. Wang, M.~D. Lukin, and E.~Demler.
\newblock {Disordered Bose-Einstein Condensates in Quasi-One-Dimensional
  Magnetic Microtraps}.
\newblock {\em Phys. Rev. Lett.}, 92:076802, 2004.
\newblock
  \href{http://journals.aps.org/prl/pdf/10.1103/PhysRevLett.92.076802}{\tt doi=
  10.1103/PhysRevLett.92.076802}.

\bibitem{PRB77-201407(R)}
Y.~Japha, O.~Entin-Wohlman, T.~David, R.~Salem, S.~Aigner, J.~Schmiedmayer, and
  R.~Folman.
\newblock {Model for organized current patterns in disordered conductors}.
\newblock {\em Phys. Rev. B}, 77:20140, 2008.
\newblock \href{http://journals.aps.org/prb/pdf/10.1103/PhysRevB.77.201407}{\tt
  doi= 10.1103/PhysRevB.77.201407}.

\bibitem{Science319-1226}
S.~Aigner, L.~Della Pietra, Y.~Japha, O.~Entin-Wohlman, T.~David, R.~Salem,
  R.~Folman, and J.~Schmiedmayer.
\newblock {Long-Range Order in Electronic Transport Through Disordered Metal
  Films}.
\newblock {\em Science}, 319:1226, 2008.
\newblock \href{http://www.sciencemag.org/content/319/5867/1226.full.pdf}{\tt
  doi= 10.1126/science.1152458}.

\bibitem{PRA75-063406}
T.~Fernholz, R.~Gerritsma, P.~Kr\"uger, and R.~J.~C. Spreeuw.
\newblock {Dynamically controlled toroidal and ring-shaped magnetic traps}.
\newblock {\em Phys. Rev. A}, 75:063406, 2007.
\newblock \href{http://journals.aps.org/pra/pdf/10.1103/PhysRevA.75.063406}{\tt
  doi= 10.1103/PhysRevA.75.063406}; see also also T. Fernholz~\etal, in
  preparation.

\bibitem{Nature415-39}
M.~Greiner, O.~Mandel, T.~Esslinger, T.~W. H\"ansch, and I.~Bloch.
\newblock {Quantum phase transition from a superfluid to a Mott insulator in a
  gas of ultracold atoms}.
\newblock {\em Nature}, 415:39, 2002.
\newblock
  \href{http://www.nature.com/nature/journal/v415/n6867/full/415039a.html}{\tt
  doi= 10.1038/415039a}.

\bibitem{JPhysB40-R61}
R.~Gati and M.~K. Oberthaler.
\newblock {A bosonic Josephson junction}.
\newblock {\em J. Phys. B}, 40:R61, 2007.
\newblock
  \href{http://iopscience.iop.org/0953-4075/40/10/R01/pdf/0953-4075_40_10_R01.pdf}{\tt
  doi= 10.1088/0953-4075/40/10/R01}.

\bibitem{IntJOpt17-857}
R.~C. Preston.
\newblock {Asymmetry in the diffraction pattern of a grating due to periodic
  errors}.
\newblock {\em Optica Acta: Int. J. Opt.}, 17:857, 1970.
\newblock \href{http://www.tandfonline.com/doi/pdf/10.1080/713818259}{\tt doi=
  DOI: 10.1080/713818259}.

\bibitem{Gamma_ratio}
{{$\Gamma_{\rm R}$ is related to the Josephson frequency ($\omega_{\rm J}$). In
  principle $\Gamma_{\rm R}\simeq\omega_{\rm J}/10$~\cite{PRA80-053619}
  and~$\Gamma_{\rm D}$ is the same as the spin-flip rate originating from
  short-correlation-length noise~\cite{ApplPhysB76-173}. The total spin-flip
  rate in our trap is~$\rm3\,Hz$. As for our specific experiment we have
  significant uncertainties concerning the values of the chemical potential,
  the tunneling rate and the correlation length of the technical noise, we do
  not quote here an estimated value for~$\Gamma_{\rm D}/\Gamma_{\rm R}$.}}

\bibitem{PRL99-060402}
Y.~Japha, O.~Arzouan, Y.~Avishai, and R.~Folman.
\newblock {Using Time-Reversal Symmetry for Sensitive Incoherent Matter-Wave
  Sagnac Interferometry}.
\newblock {\em Phys. Rev. Lett.}, 99:060402, 2007.
\newblock
  \href{http://journals.aps.org/prl/pdf/10.1103/PhysRevLett.99.060402}{\tt doi=
  10.1103/PhysRevLett.99.060402}.

\bibitem{EurPhysJD48-321}
T.~David, Y.~Japha, V.~Dikovsky, R.~Salem, C.~Henkel, and R.~Folman.
\newblock {Magnetic interactions of cold atoms with anisotropic conductors}.
\newblock {\em Eur. Phys. J. D}, 48:321, 2008.
\newblock
  \href{http://link.springer.com/article/10.1140%2Fepjd%2Fe2008-00119-x#page-1}{\tt
  doi= 10.1140/epjd/e2008-00119-x}.

\bibitem{PRA79-043403}
P.~G. Petrov, S.~Machluf, S.~Younis, R.~Macaluso, T.~David, B.~Hadad, Y.~Japha,
  M.~Keil, E.~Joselevich, and R.~Folman.
\newblock {Trapping cold atoms using surface-grown carbon nanotubes}.
\newblock {\em Phys. Rev. A}, 79:043403, 2009.
\newblock \href{http://journals.aps.org/pra/pdf/10.1103/PhysRevA.79.043403}{\tt
  doi= 10.1103/PhysRevA.79.043403}.

\bibitem{PRA80-053619}
E.~Boukobza, D.~Cohen, and A.~Vardi.
\newblock {Interaction-induced dynamical phase locking of Bose-Einstein
  condensates}.
\newblock {\em Phys. Rev. A}, 80:053619, 2009.
\newblock \href{http://journals.aps.org/pra/pdf/10.1103/PhysRevA.80.053619}{\tt
  doi= 10.1103/PhysRevA.80.053619}.

\end{thebibliography}

\end{document}